# Magnetohydrodynamic simulation of coronal mass ejections using interplanetary scintillation data observed from radio sites ISEE and LOFAR

Kazumasa Iwai[a] Richard A. Fallows[b, c], Mario M. Bisi[c], Daikou Shiota[d,a], Bernard V. Jackson[e], Munetoshi Tokumaru[a], Ken'ichi Fujiki[a]

a. Institute for Space-Earth Environmental Research, Nagoya University, Furo-cho, Chikusa-ku, Nagoya, 464-8601, Japan
b. ASTRON - The Netherlands Institute for Radio Astronomy, Oude Hoogeveensedijk 4, 7991 PD, Dwingeloo, The Netherlands)
c. RAL Space, United Kingdom Research and Innovation UKRI) - Science & Technology Facilities Council (STFC) - Rutherford Appleton Laboratory (RAL), Harwell Campus, Oxfordshire, OC11 0QX, UK
d. National Institute of Information and Communications Technology, 4-2-1 Nukui-kita, Koganei, Tokyo 184-8795, Japan
e. Center for Astrophysics and Space Sciences, University of California, San Diego, LaJolla, California, 92093-0424, USA

**Abstract**

Interplanetary scintillation (IPS) is a useful tool for detecting coronal mass ejections (CMEs) throughout interplanetary space. Global magnetohydrodynamic (MHD) simulations of the heliosphere, which are usually used to predict the arrival and geo-effectiveness of CMEs, can be improved using IPS data. In this study, we demonstrate an MHD simulation that includes IPS data from multiple stations to improve CME modelling. The CMEs, which occurred on 09-10 September 2017, were observed over the period 10-12 September 2017 using the Low-Frequency Array (LOFAR) and IPS array of the Institute for Space-Earth Environmental Research (ISEE), Nagoya University, as they tracked through the inner heliosphere. We simulated CME propagation using a global MHD simulation, SUSANOO-CME, in which CMEs were modeled as spheromaks, and the IPS data were synthesised from the simulation results. The MHD simulation suggests that the CMEs merged in interplanetary space, forming complicated IPS g-level distributions in the sky map. We found that the MHD simulation that best fits both LOFAR and ISEE data provided a better reconstruction of the CMEs and a better forecast of their arrival at Earth than from measurements when these simulations were fit from the ISEE site alone. More IPS data observed from multiple stations at different local times in this study can help reconstruct the global structure of the CME, thus improving and evaluating the CME modelling.

## 1 Introduction

Solar eruptive phenomena generate coronal mass ejections (CMEs) that propagate into interplanetary space and sometimes arrive at Earth. CMEs cause various disturbances around Earth, where many societal infrastructures are in operation. Currently, CMEs have become an important target of space weather studies (e.g. Temmer et al., in prep).

Interplanetary scintillation (IPS) is a radio-scattering phenomenon caused by density irregularities in the solar wind (Clarke, 1964; Hewish et al., 1964). Ground-based radio telescopes that observe extragalactic radio sources, such as quasars, can detect radio scintillation caused by the outflow of solar plasma throughout the inner heliosphere between the radio telescope and distant, compact radio sources. A fast-propagating CME can sweep the background solar wind, and a high-density region is formed in front of the CME. This region can significantly scatter radio emissions that can be detected by observations of IPS. Observations of IPS have thus been used to detect CMEs propagating in interplanetary space (e.g. Tokumaru et al., 2003; Manoharan, 2006; Bisi et al., 2010).

Most IPS-dedicated ground-based radio telescopes observe around their local meridians (e.g. Tokumaru et al., 2011; Chashei et al., 2013; Gonzalez-Esparza et al., 2021), which is equivalent to scanning the inner heliosphere once per day. Daily meridian scan observations have been used to reconstruct the global structure of solar wind using tomography techniques (e.g. Jackson et al., 1998; Kojima et al., 1998; Jackson et al., 2020). On the other hand, fast-propagating CMEs sometimes arrive at Earth within a few days, which requires higher cadence observations of IPS. The coordinated observation of multiple IPS stations located at different longitudes can improve the cadence of IPS observations and the ability to improve tomographic reconstructions of the inner heliosphere (e.g., Tokumaru et al., 2019; Jackson et al., 2022). This idea is known as the Worldwide IPS Stations (WIPSS; Bisi et al., 2017) Network, and several observational projects have been involved. The observation of IPS, by tracking a specific radio source, can detect the exact crossing time of the CME front at the line-of-sight (LOS) of the tracked radio source, measuring the spatial variation of velocity and density irregularities along the propagation direction of the CME (Gothoskar et al., 1999). A recent array such as the Low-Frequency Array (LOFAR; van Haarlem et al., 2013) with a high sensitivity, wide field of view, and steerable beams, could be used for this purpose, as described in Fallows *et al.* 2023 (this issue).

Global magnetohydrodynamic (MHD) simulations, such as ENLIL (Odstrcil, 2003), EUHFORIA (Pomoell and Poedts, 2018), and SUSANOO-CME (Shiota et al., 2014; Shiota and Kataoka, 2016), are widely used to reconstruct CME propagation and forecast their arrival in space-weather research and operations. In typical simulations, a CME is assumed to be a cone or spheromak at the inner boundary around 20–30 Rs and this is injected into the background solar wind, providing a simulated CME propagation into interplanetary space. However, prediction of the geo-effectiveness of CMEs is still challenging (e.g., Vourlidas et al. 2019). These MHD simulations have typically an error of more than 10 h in Time of Arrival (ToA) (Riley et al. 2018; Wold et al. 2018). Recent work by Iwai et al. (2019) and (2021) enabled the calculation of IPS indices from the global density distribution derived from the SUSANOO-CME MHD simulation. These studies, along with that by Bisi et al. (2009) using Ooty IPS data, suggest that the more IPS data are included in the tomographic reconstruction and/or MHD simulation, the better the reconstruction and forecasts achieved.

In this study, we demonstrate the coordinated observations of IPS using the LOFAR system and the Institute for Space-Earth Environmental Research (ISEE) arrays and include their data in the SUSANOO-CME MHD simulation to improve its accuracy.

## 2 Method
### 2.1 Observations of IPS by ISEE

Observations of IPS by ISEE, Nagoya University, were carried out by the Solar Wind Imaging Facility (SWIFT; Tokumaru et al., 2011), which is one of the three large IPS radio telescopes of ISEE. The observing frequency of this telescope is 327 MHz, which enables the detection of IPS signatures from the solar wind between $\sim 0.2$ AU and 1.0 AU. This telescope has a cylindrical parabolic antenna fixed in the North-South direction. The phased-array receiver system of this telescope has a single beam that is steerable along the meridian line. Approximately 50–70 radio sources are observed sequentially at the time of their meridian transit between the morning and evening of each day.

The amplitude of the scintillation is derived from each radio source from the IPS observation. The amplitude of the scintillation is primarily determined by the electron density irregularities along the LOS to the radio source. However, this information also contains information about other effects such as radio source elongation and radio source size. The ratio between the observed scintillation amplitude and typical scintillation amplitude of the radio source at given elongation is defined as a g-level (Gapper et al 1982). The typical scintillation amplitude is derived by a fitting of the daily scintillation amplitude (e.g. Fig. 3 of Jackson et al 2022). Using g-level measurements, elongation and radio source size effects are mostly eliminated; an enhancement of g-level also suggests that there are some high density irregularity regions along the LOS.

### 2.2 Observations of IPS using LOFAR

LOFAR is a radio interferometer located in Europe. The flat array system and fully steerable phased-array beams enable tracking of specific radio sources for a long time each day. The observations of IPS and subsequent data analysis of LOFAR are described in more detail in Fallows et al (2020) and a companion paper by Fallows et al. 2023 (this issue). LOFAR observed a radio source 3C147 on 12 September 2017 and detected its IPS response. The elongation and position angle of this radio source was 82° and 297°, respectively. The observation frequency range was 110–190 MHz with an integration time of 0.01 s.

### 2.3 MHD simulation with SUSANOO

The global MHD simulation SUSANOO-CME enables the reconstruction of three-dimensional (3-D) global structure of the heliosphere and its variation with time. The details of this simulation system have been described in previous studies (Shiota et al., 2014; Shiota and Kataoka, 2016). The inner-heliosphere region is simulated in spherical coordinates between 25 and 425 Rs, formed by a Yin-Yang grid (Kageyama and Sato, 2004). In this way, the North and South polar regions of the heliosphere are filled with grid points enabling us to calculate IPS in the northern and southern regions in each simulation step. The magnetic field of the inner boundary is given by the potential field source surface (PFSS) approximation of the photospheric magnetic field (Schatten et al., 1969). In this model, the three-dimensional coronal magnetic field is reconstructed by assuming that the corona is in a current free condition and all the field lines open out at the source surface that is fixed at 2.5 Rs in this study. The magnetic field on the inner boundary of the simulation (25Rs) is obtained by extending the radial magnetic field on the source surface assuming conservation of the open magnetic flux (Shiota et al, 2014). The velocity, density, and temperature of the background solar wind are given by empirical models of solar wind (Wang and Sheeley, 1990; Arge and Pizzo, 2000; Hayashi et al., 2003). The CME, approximated as a spheromak, was placed on the inner boundary. The spheromak has ten free parameters that express the onset time, location (longitude and latitude), size (radial and angular), radial velocity, and magnetic characteristics (flux, chirality, tilt angle, and inclination angle) at the inner boundary.

IPS indices were calculated from the density distribution of the SUSANOO-CME simulations. In this study, we assume that the density fluctuation in the solar wind is proportional to the density of the solar wind, and that the weak scattering condition is valid in the simulation region. We can then calculate the amount of radio scintillation, the so-called m-index, along the LOS of the radio sources by integrating the density convolved with the weighting function of the IPS (Young, 1971; Iwai et al., 2019). First, we simulate only the background solar wind and derive the IPS m-index of the solar wind. We then simulate the solar wind with the spheromak and derive the IPS m-index. The ratio between the simulated IPS indices with and without the spheromak can be recognised as the IPS g-level, which is compared with the observations. The proportionality coefficient between the density and density fluctuation, which is unknown in the MHD simulation, is cancelled out in the calculated g-level that is a ratio between the two m-indices. Details of the IPS calculation are provided in Iwai et al. (2019; 2021).

### 3 Result

#### *3.1 CME event on 10 September 2017*

In September 2017, a large active region, NOAA 12673, generated numerous solar flares. This active region was located on the western limb (S09W91) on 10 September 2017. There were several CMEs in this active region between 9 and 10 September 2017. We consider three of them that were clearly observed by the Large Angle and Spectrometric Coronagraph (LASCO: Brueckner et al., 1995) C3 imager onboard the Solar and Heliospheric Observatory (SOHO). The first CME (CME1) was generated by a C-class flare that started at 14:50 UT on 9 September 2017 (Figure 1a). The linear speed of this CME is very slow (473 km/s in the SOHO LASCO CME catalog; Yashiro et al., 2004). An M-class flare at around 22:00 UT on 9 September 2017 erupted a second CME (CME2) that became a partial halo CME in the LACSO field of view (Figure 1b). The linear speed of CME2 was approximately 1,000 km/s, which was faster than that of CME1, and they merged in the C3 field of view (Figure 1c). An X8.2 flare, started at 15:35 UT on 10 September 2017, generated a full halo CME with an extremely fast initial speed (Figure 1d). The initial speed archived in the LASCO CME catalog was greater than 3,000 km/s. The linear speed of these CMEs are calculated through the white light images. This velocity is a projection on the plane-of-sky of the true radial propagation speed and is underestimated (see e.g. Burkepile et al., 2004; Vrsnak et al., 2007; Temmer et al., 2009). However, as the active region was located on the western limb, the projection effects are negligible (e.g. Paouris et al., 2021). Therefore, the linear speed should be close to the propagation speed. This CME generated a severe solar energetic particle (SEP) event and produced also a Ground Level Enhancement (GLE) event (Gopalswamy et al., 2018; Kouloumvakos et al., 2020; Mavromichalaki et al., 2018). Hereafter, this CME is referred to as CME3. The disturbance caused by the CMEs arrived at Earth on 12 September 2017. The shock-arrival time observed by the Wind spacecraft in orbit about the Sun-Earth L1 point was approximately 19:30 UT on 12 September 2017, with a maximum speed of approximately 600 km/s. The details of these flares and related space-weather events have been reported in many studies (e.g. Warren et al., 2018; Gopalswamy et al., 2018; Veronig et al., 2018; Lee et al., 2018).

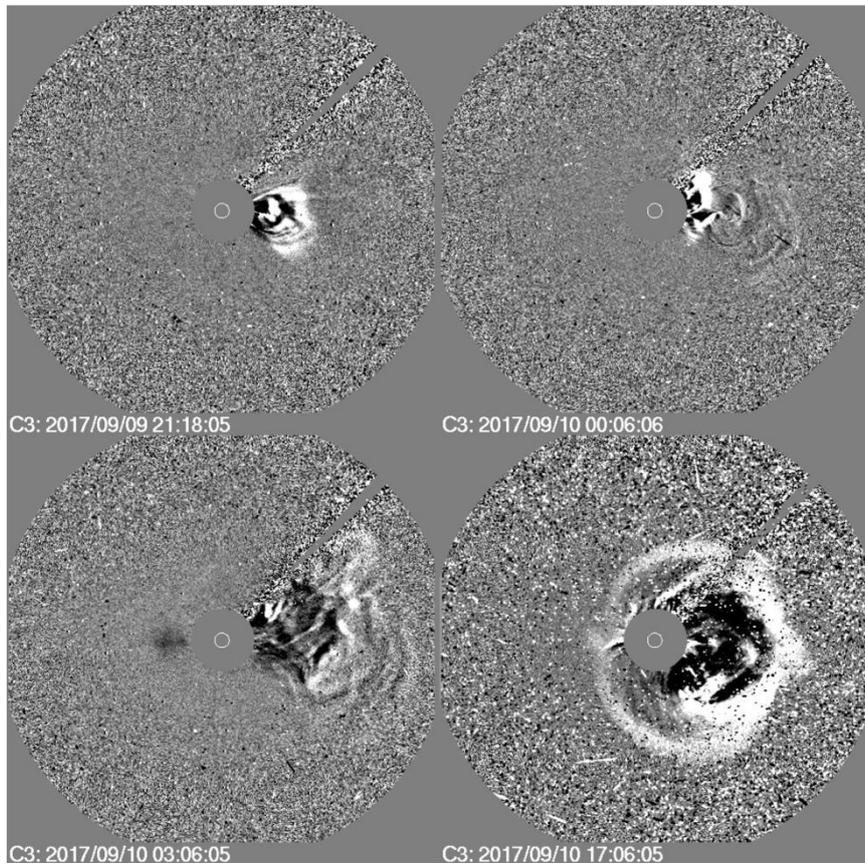

Figure 1 Coronagraph difference images obtained from Solar and Heliospheric Observatory (SOHO) Large Angle and Spectrometric Coronagraph (LASCO) C3 at time of (a) CME1, (b) CME2, (c) merger of CME1 and CME2, and (d) CME3.

### 3.2 Observations of IPS using both ISEE and LOFAR radio-telescope systems

Figure 2 shows the IPS g-level observed by ISEE between 11 and 13 September 2017, which was projected onto the sky map. The X- and Y-axis indicate elongation from the Sun in radians. The observations of IPS times for each radio source corresponds to its local meridian passage time. Therefore, the observation of Figure 2 started from the West edge at around 21:00 UT and scanned toward the East side until 09:00 UT of the next day. The radio source 3C147, tracked by LOFAR on 12 September 2017 was also observed by ISEE. The location of 3C147 is indicated by the arrows in Figure 2. They show no significant enhanced IPS response was derived from that radio source. For fitting CME models to observations there is significant advantage in as near continuous monitoring of radio sources as possible. Figure 3 shows the IPS g-level of 3C147 as observed by LOFAR from 22:00 UT on 11 September 2017 to 12:00 UT on 12 September 2017. This figure shows that there was an impulsive g-level enhancement on 12 September 2017 between the ISEE daily observations.

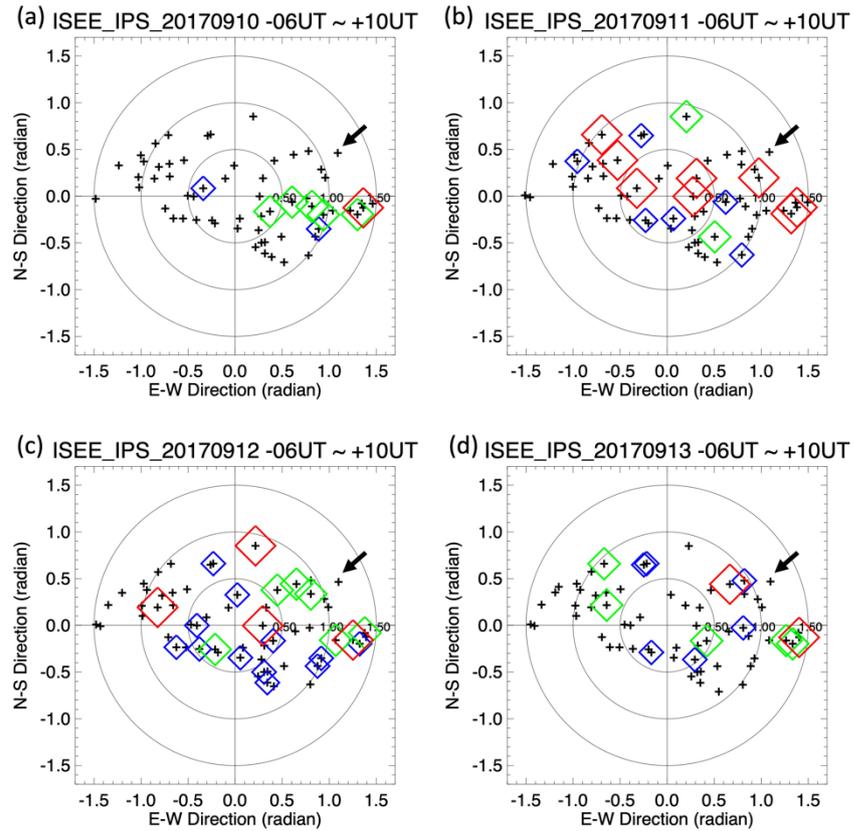

Figure 2 IPS g-level projection on the sky map in the heliographic coordinates obtained from ISEE on (a) 10, (b) 11, (c)12, and (d) 13 September 2017. X- and Y-axis indicate elongation from the Sun in radians. +: all observed radio sources, diamonds: radio sources with g-level values as follows: $1.2 < g < 1.5$ (blue) $1.5 < g < 2.0$ (green), and $2.0 < g$ (red). The location of 3C147 as observed by LOFAR is indicated by the black arrow.

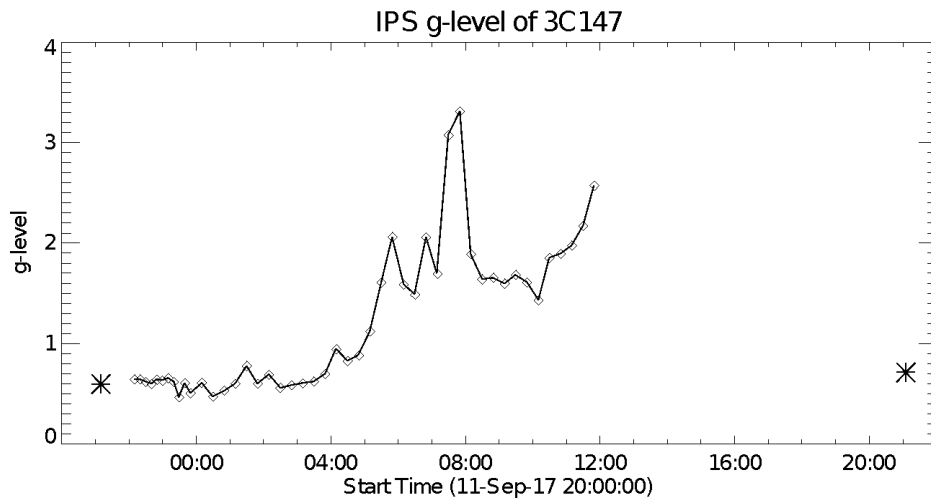

Figure 3 IPS g-level of 3C147 on September 12 as observed by LOFAR. The IPS g-level of 3C147 observed by ISEE before and after the LOFAR observation is also shown in asterisks.

### 3.3 MHD simulation by SUSANOO-CME

Figure 4 shows an example of the SUSANOO-CME MHD simulation results as a background colour and observed IPS data as symbols, projected onto a sky map. The background colour of the left panel indicates the simulated IPS m-index. The simulated m-index becomes larger in the region closer to the Sun because the simulation assumes that the amount of density irregularity should be proportional to the density itself, and the density decreases with the square of the distance from the Sun. There are also high m-index regions (i.e high density along the LOS) at lower latitudes, corresponding to the heliospheric current sheet (HCS). The right panel shows the IPS g-levels derived from the MHD simulation. The simulation assumed that the m-index ratio with and without the CME is the g-level value of the IPS. Therefore, the radial variation and HCS disappeared from the g-level maps. The high-g-level regions exhibited loop-like distributions. This is because the faster-propagating CMEs sweep the background solar wind to form a high-density spherical shell in front of the CME.

In this simulation, CME1 and CME2 had already merged before they reached the inner boundary of the simulation. The merger appeared in the simulation region around 04:00 UT on 10 September 2017. CME3 then reached the merger around 00:00 UT on 11 September 2017. The simulated g-level was enhanced by the CME-CME interaction. This may correspond to the strong g-levels observed at 01:00 UT on 11 September 2017. After the CME-CME interaction, a complicated merger was formed and propagated through the inner heliosphere. We presume CME3 was decelerated by both the background solar wind and merger of CME1 and CME2. Therefore, the merger of CME1, CME2, and CME3 had a greatly reduced speed when it arrived at the Earth.

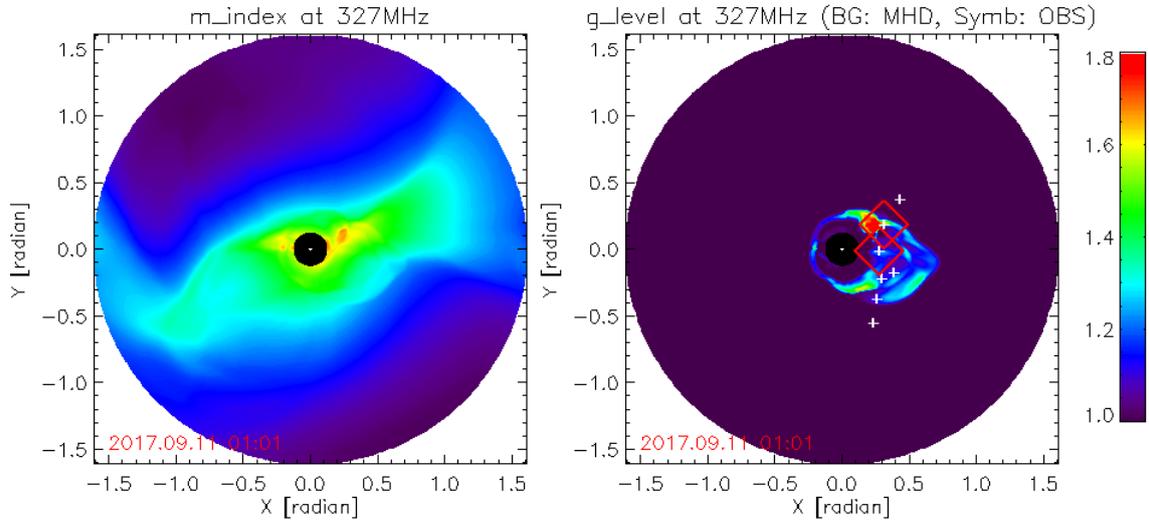

Figure 4 Simulated IPS indices obtained from SUSANOO-CME MHD simulation RUN 7 at 01:00 UT on 11 September 2017. Left: Amplitude of the IPS, m-index. Right: the IPS ratio between simulations with and without spheromaks, g-level. The IPS data observed between 01:00 UT and 02:00 UT are superposed on the right panel in the same format as that of Figure 2. The corresponding gif movie is available as supplementary material.

### 3.4 Difference of the CME parameters and comparison with the observations of IPS

The initial parameters of the spheromak were defined from the Geostationary Operational Environmental Satellite (GOES) and LASCO observational data in the same manner as in Iwai et al. (2021). The spheromak parameters of CME1 and CME2 were fixed because these CMEs merged in the early phase of the propagation, and we could not fit their parameters using the IPS observations. The parameters of the corresponding spheromaks are presented in Table 1.

Table 1 Parameters of the spheromak included in the MHD simulation

| CME | Onset time | Longitude | Latitude | Velocity | Radial width | Angular width | $B$ (Mx)[a] |
|-----|-----------|-----------|----------|----------|-------------|---------------|-------------|
| 1 | 09.09 14:49 | 88° | −9° | 473 km/s | 2 Rs | 30° | 0.3E+21 |
| 2 | 09.09 22:04 | 88° | −9° | 1019 km/s | 3 Rs | 60° | 1.0E+21 |
| 3 | 09.10 15:35 | 88° | −8° | 1500–2400 km/s | 4 Rs | 90°–180° | 3.0 E+21 |

[a] total magnetic flux contained in the spheromak (Mx)

In this study, the free parameters of CME3 were the velocity and angular width of the spheromak. Iwai et al. (2021) showed that the initial velocity is the simplest parameter for observing the difference in the IPS g-level distribution and evaluating the time of arrival (ToA) of the CME at Earth. In addition, we investigated the change in angular width because the CMEs of interest in this study were launched from the limb. Therefore, the angular width determines the earthward structure, as discussed in the next section. To find the appropriate range of the free parameters, we tested simulations that have different initial velocities between 1000 km/s and 3000 km/s. We found that the initial velocity between 1500 km/s and 2400 km/s with appropriate angular width can reconstruct the ToA at the Earth. The CME parameters of each simulation run shown in the following figures are summarized in Table 2.

Table 2 Parameters of spheromak for CME3 in each simulation run.

| RUN | Velocity | Velocity step | Width | Width step |
|-----|----------|---------------|-------|------------|
| 1 to 10 | 2,000 km/s | --- | 90°–180° | 10° |
| 11 to 20 | 1,500–2,400 km/s | 100 km/s | 135° | --- |

Comparison between the simulated and observed IPS follows the method developed by Iwai et al. (2019; 2021). This system compares the location of the simulated and observed high-g-level regions, that are usually formed at the front part of the CME, at given times with a 1 h cadence. There might be some high density regions behind the CME front. Therefore, this system only considers the front of the high-g-level regions along specific declination angles from the solar north pole to avoid such regions (Iwai et al., 2019).

First, we fixed the initial velocity of 2,000 km/s, which is close to the maximum speed derived from the computer-aided CME tracking software (CACTus: Robbrecht and Berghmans 2004; http://sidc.be/cactus/), using the LASCO data and changed the radial width between 90° and 180° with 10° steps (RUN 1 to 10). Figure 5 shows the g-level distribution of RUN 1, 4, 7, and 10 superimposed on the observed g-level on 12 September 2017. The simulation that best fit the IPS among RUN 1 to 10 was RUN 7. Figure 7a shows the time variation of the solar wind velocity at the Earth's location derived from each simulation. Any simulation that uses a smaller angular width of the spheromak does not show a clear CME arrival at Earth such as RUN1 and RUN2 that are shown in black and purple lines, respectively.

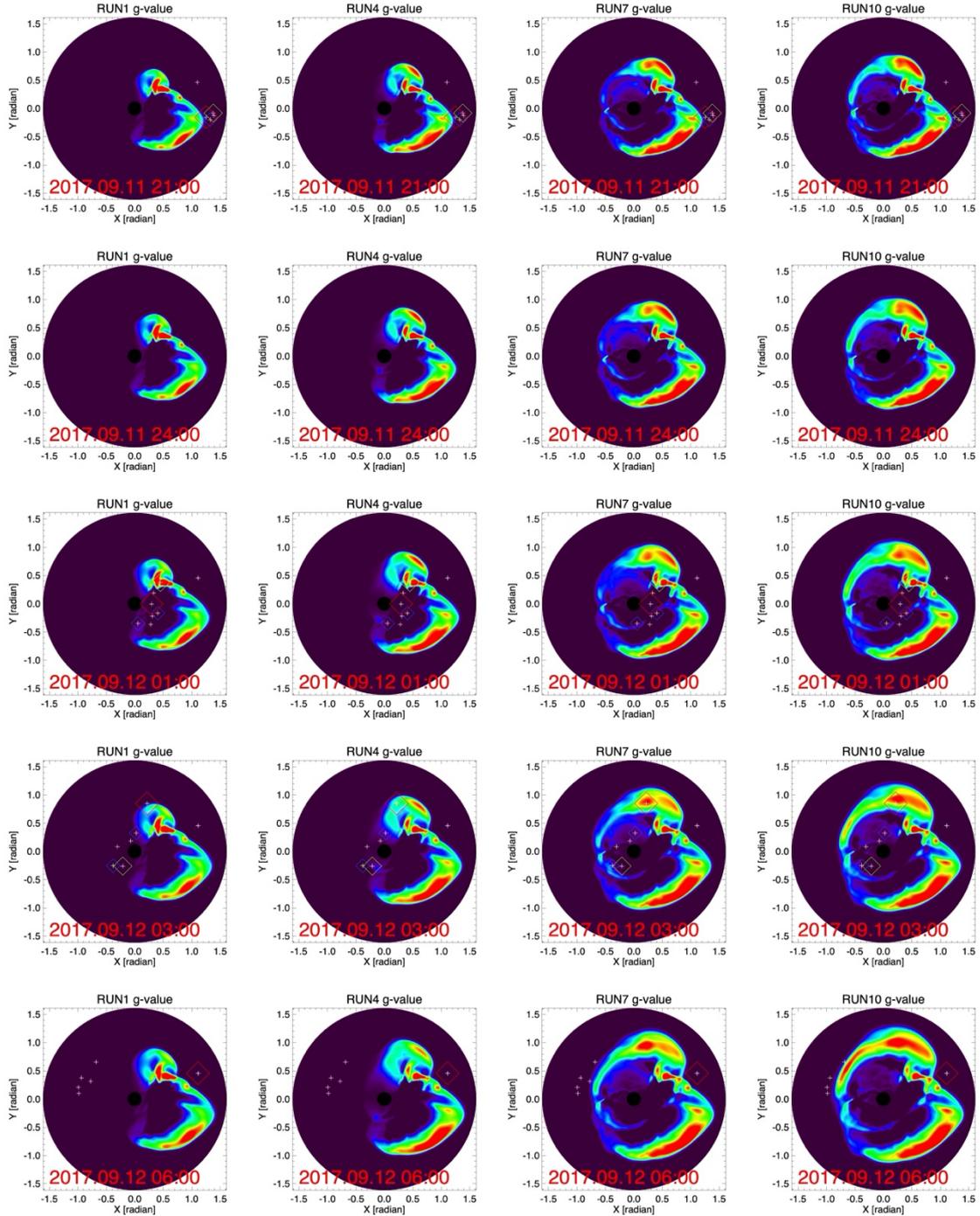

Figure 5 Comparison between simulated (the background) and observed (symbols) IPS g-levels on 11-12 September 2017. Simulations (columns from the left): RUN1, RUN4, RUN7 and RUN10. Time (rows from the top: 21:00 UT, 23:00 UT, 1:00 UT, 3:00 UT, and 5:00 UT).

Next, we fixed the initial radial width of 150°, which gives the best IPS estimation at 1 AU with an initial velocity of 2,000 km/s, and changed the initial velocity between 1,500 km/s and 2,400 km/s. Figure 6 shows the g-level distribution of RUN 11, 14, 17, and 20 superimposed on the observed g-level on 12 September 2017. The best fit to IPS among RUN 11 to 20 was derived from RUN 15 (velocity of 1,900 km/s). Figure 7b shows the time variation of the solar wind velocity at the Earth's location derived from each simulation. RUN 7 fits slightly better than RUN 15. The simulation run that best fits the IPS data (RUN 7) has about 3.6 h difference from the observed shock arrival.

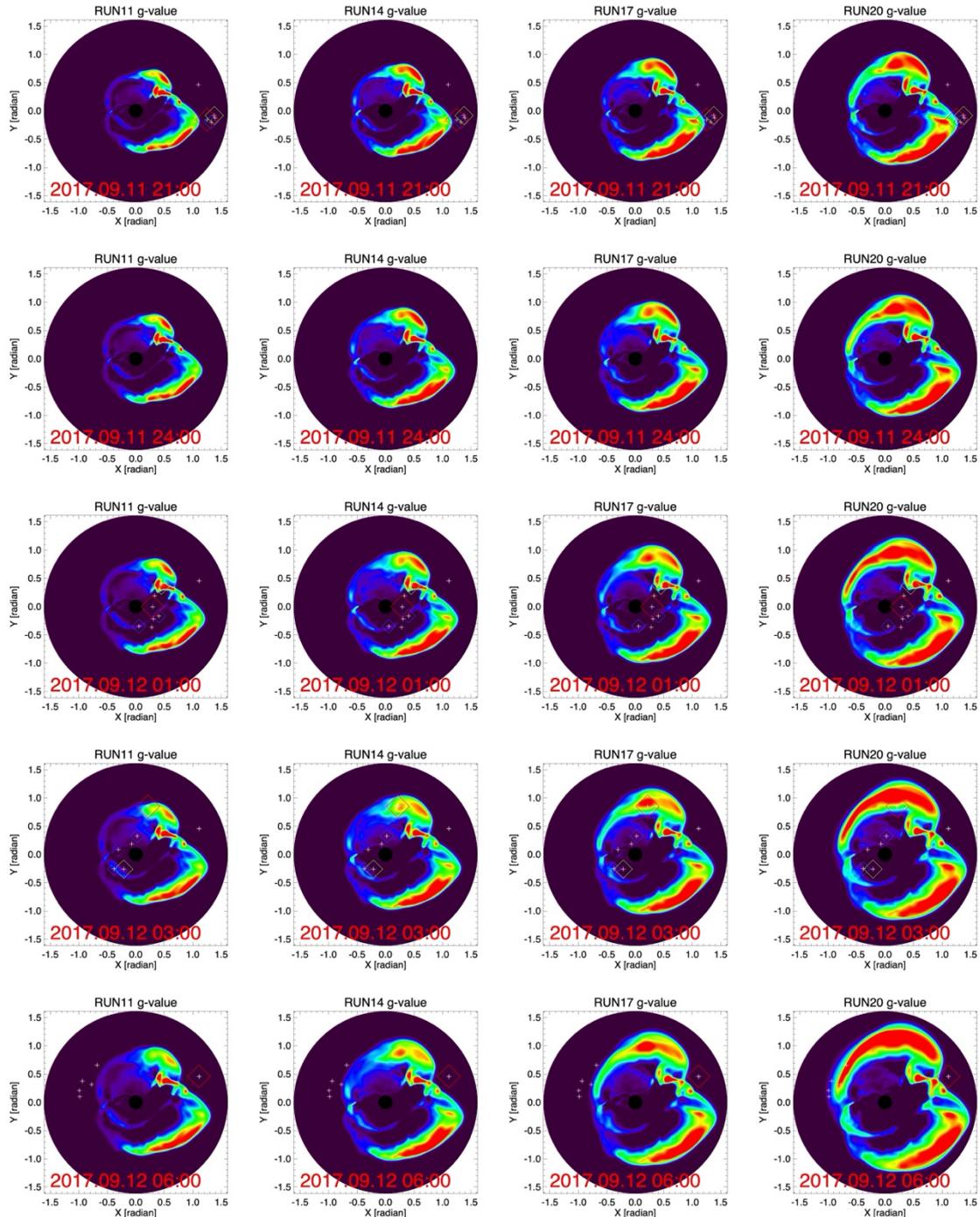

Figure 6 Comparison between simulated (the background) and observed (symbols) IPS g-levels on 11 September 2017. Simulations (columns from the left): RUN11, RUN14, RUN17 and RUN20. Time (rows from the top: 21:00 UT, 23:00 UT, 1:00 UT, 3:00 UT, and 5:00 UT).

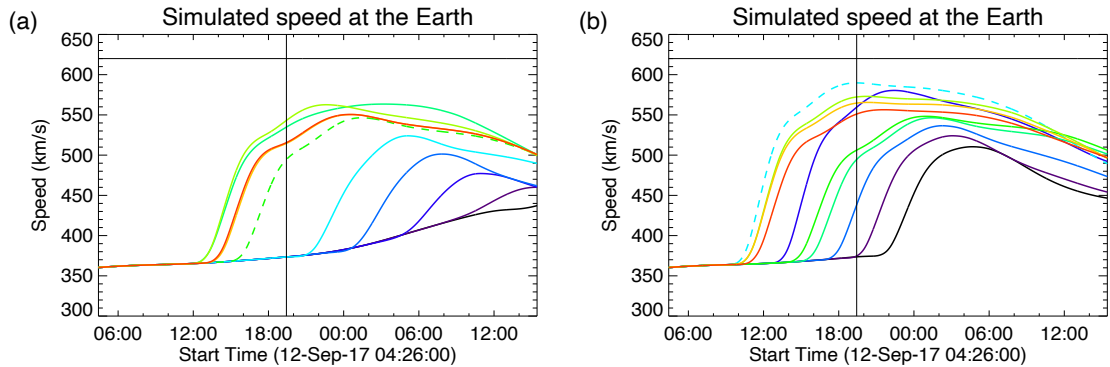

Figure 7 Solar wind speed at Earth derived by simulation runs with CMEs with different initial parameters; (a) RUN 1–10, (b) RUN 11–20. The line colors correspond to the RUN from black (1 or 11) to red (10 or 20). Vertical line: arrival time at the Earth. Horizontal line: typical speed of the CME measured by Wind. Dashed lines indicate the best fit to the IPS of each panel; these are RUN 7 and RUN 15 for figure panel (a) and (b), respectively.

## 4 Discussion

### 4.1 Accuracy of the CME reconstruction by using increased IPS data

It is difficult to define the predicted ToA without the IPS data of this event because the typical initial parameters used in the previous SUSANOO studies, for example, 90° for the angular width, with the maximum speed of CACTus as the initial velocity cannot predict even the arrival of the CME itself. Now, we know that CME3 arrives at Earth, which requires a wider angular width. If we use CACTus' initial velocity, 2,000 km/s, and angular width as a free parameter, the best fit to IPS was derived with an angular width of 150°. This is because a simulation with a smaller angular width cannot reconstruct the observed g-level distribution to the East, resulting in a lower score in the comparison sequence between the observed and simulated IPS data. This is most clearly shown in Figure 5 at 05:00 UT, when the ISEE array observed the eastern portion of the sky and LOFAR observed the western part of the sky simultaneously. Note that ISEE also observed 3C147 on 11 September 2017 at 21:00 UT, when the CME had not arrived at the 3C147 LOS, before it was possible to distinguish the halo structure in the IPS sky map.

It should also be mentioned that the MHD simulation that based only on the ISEE data cannot predict the arrival of the CME to the Earth because the ISEE data cannot find the earthward component of the CME. The standalone LOFAR IPS data with MHD simulation also cannot find the westward component. The halo distribution of the IPS, recognised only by the combined ISEE and LOFAR IPS datasets, suggests that the CME has an earthward component. This observed IPS distribution is consistent with the white-light coronagraph observations, which also showed a halo structure (Figure 1d) and corresponding shock arrival at Earth. Simulation runs reconstructing the IPS halo structure predict arrival of the CME at Earth (Figure 7a). This result suggests that increased numbers of IPS observations would be of great help to reconstruct the global structure of this CME.

### 4.2 Time variation of IPS at the 3C147 location

The tracking observation of 3C147 by LOFAR should determine the exact onset time of g-level enhancement. However, in this study, the simulation still had a time delay in the g-level enhancement from the LOFAR observation. A HCS was observed on the trajectory of the CMEs to 3C147 (left panel of Figure 4). Figure 8 shows the density distribution on the ecliptic plane derived from the SUSANOO simulations. The interaction between the HCS and CMEs caused deformation of the CMEs in our simulation. This deformed region corresponds to the LOS of 3C147. A possible explanation is that the actual location of the HSC is slightly different from that in the simulation, which may cause different deformations along the 3C147 trajectory.

The magnetic field at the inner boundary of our simulation is given by the PFSS approximation using the GONG magnetogram with a source surface height of 2.5 Rs. The difference between these assumptions and the real conditions can result in different HCS locations. The magnetogram can be improved by a flux transport model, such as the air force data assimilative photospheric flux transport (ADAPT) model (Arge et al., 2010). The source surface height can also be modified by coronal hole observations or in situ measurements of the inner heliosphere (e.g. Badman et al., 2020). The background solar wind provided by the empirical model in this study was also improved by observations of IPS via the tomography technique (e.g. Jackson et al., 2015; 2023 - this issue). These possible future improvements in SUSANOO will provide a better reconstruction of the HCS.

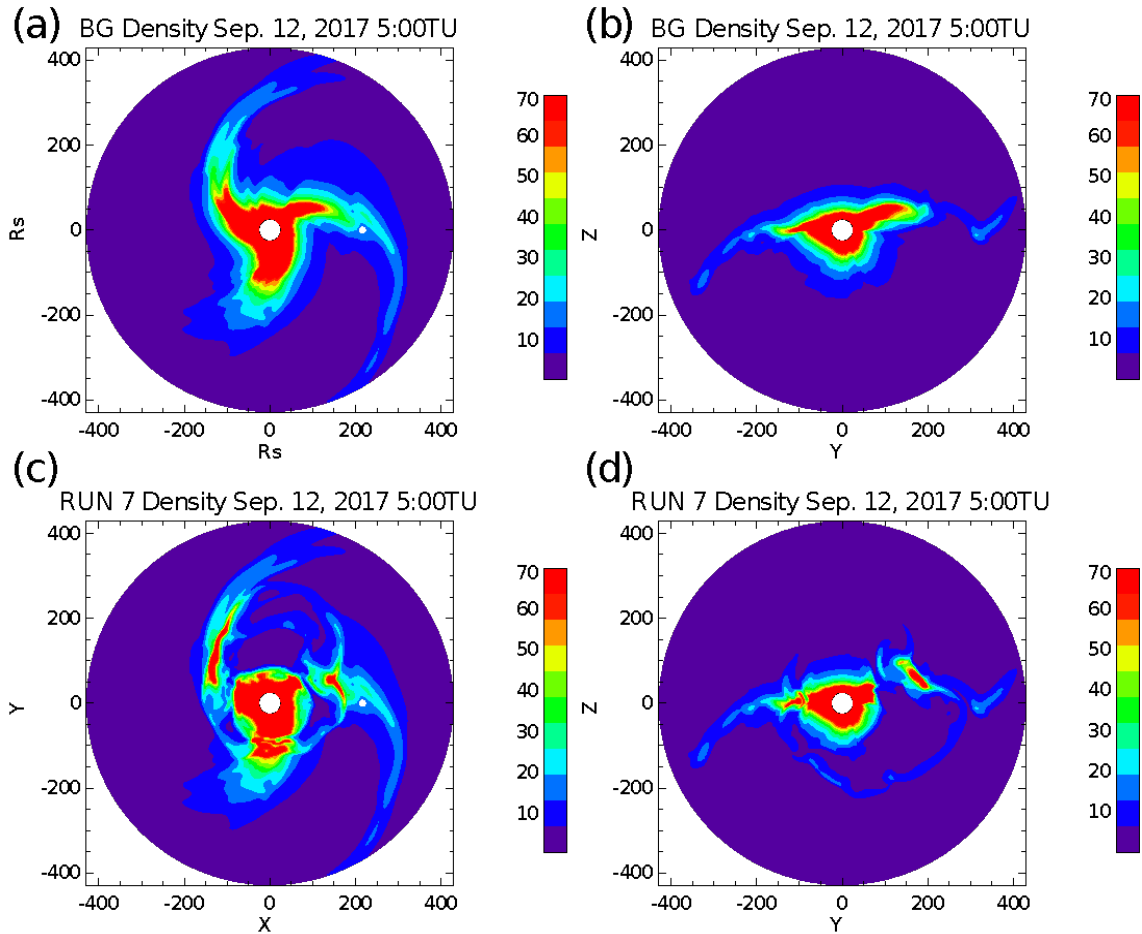

Figure 8 Density distribution in the ecliptic plane derived from SUSANOO (a) without CME and (c) with CMEs RUN 7. Density distribution on the plane vertical to the ecliptic plane and Sun-earth line derived from SUSANOO (b) without CME and (d) with CMEs RUN 7. White circles at [216, 0] in the left panels indicate the location of Earth.

Figure 9 shows the time variation of the g-level determined and simulated at the 3C147 LOS. The black line indicates the g-level along the Sun to 3C147. The onset time of g-level is not reconstructed. Note that the amplitude of g-level cannot be reconstructed from our simulation system even if we have a perfect CME parameter values (Iwai et al., 2019; 2021). Thus we only use the simulated onset times compared with the observations to provide a fit. The red line indicates the g-level along a position angle that is inclined 30° from that of 3C147 to avoid the HCS region. Although the onset time still had a delay of a few hours, the duration of the g-level enhancement was in agreement between the observed and simulated data. Therefore, an explanation that the radial structure of the CME3 with a slightly different HCS location corresponds to the observed time variation in the g-level of 3C147 is not so strange.

Another reason for the timing difference of our reconstruction of CME3 could come from CME2. If the actual initial speed of CME2 was faster than we assumed, the deceleration of CME3 due to the interaction with CME2 would be smaller. This means CME3 would have arrived at the 3C147 location faster than we simulated as observed by LOFAR. This case corresponds to the blue line of Figure 9 that shows the time variation of g-level at a 53° (30° closer to the Sun than that of 3C147) in elongation angle, and at a 293° position angle (the same as that of the 3C147). Here, the interaction between CME2 and CME3 had not been completed, and a two-step g-level enhancement was predicted as observed by LOFAR. If the initial speed of CME2 was faster than we expected, this structure would have been simulated at the location of 3C147.

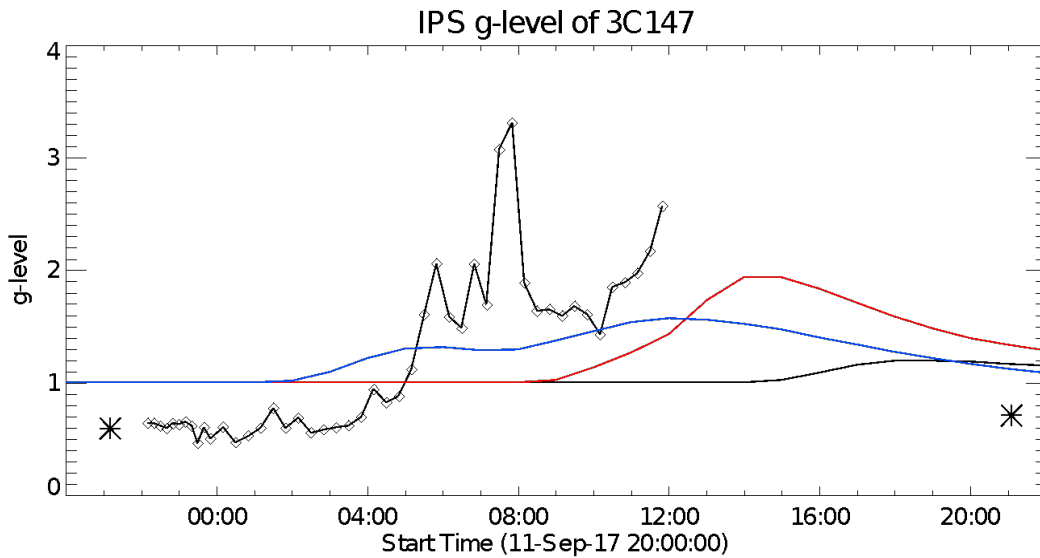

Figure 9 Time variation of the observed (symbols) and simulated by RUN 7 (lines) IPS g-levels. Black: LOFAR measurements of 3C147. Red: MHD simulated measurements at the same elongation and at a 30° inclined position angle from that of 3C147. Blue: MHD simulated measurements at the same position angle and at a 30° elongation closer to the Sun than that of 3C147.

### 4.3 Limitation of the current simulation and observations

From the point of view of a real-time space-weather forecast, it would have been difficult to improve the ToA accuracy of the 10 September 2017 CME using only the currently available IPS data. A limb CME requires the reconstruction of its 3-D structure more accurately, while a disc center event only requires an estimate of its radial components. In addition, the extremely-high initial speed allowed the radio telescope to make only a limited number of observations before its arrival. Moreover, it took several hours for the high-g-level regions to pass through a specific radio source (see Figure 9). Therefore, all-sky observations can only detect the high-g-level region with an accuracy of several hours.

We usually choose one free parameter among the ten spheromak parameters. This means that the other nine parameters are not tuned. The magnetic flux assumed in this study can be determined by the dimming of the extreme ultraviolet observations (e.g. Dissauer et al., 2019). The graduated cylindrical shell model fitting of the white-light coronagraph observation (e.g. Hess and Zhang, 2014) can also be used to derive other spheromak parameters. Further data combined with the observations of IPS in the MHD simulation can certainly help in the future to provide better reconstructions.

## 5 Summary

In this study, we analysed data from the coordinated observations of IPS using both the ISEE and LOFAR radio-telescope systems. These observation facilities are located at different longitudes, with different observation sequences and frequencies. Such different IPS data were merged in the global MHD simulation, SUSANOO-CME. The combination of multiple observations of IPS and simulations showed that with more IPS data included in the MHD simulation, a better reconstruction could be achieved. Our results can be summarised as follows:

- Our global MHD simulation reconstructed the spatial and time variation of the IPS g-levels;
- The global structure of the CME sequence is better understood by using the combined ISEE and LOFAR IPS data compared to using only the ISEE data; and
- The time variation of the g-level of 3C147 can be explained by the passage of the CME front, although there are several hours of time difference between the observed and simulated g-level onset times - this time difference can be explained by the deformation of the CME caused by the interactions between the HCS and CME.

Our study suggests that combining IPS data from different locations in the MHD simulation helps with improving CME reconstructions. The cooperation of world IPS observation stations, noted as the WIPSS Network, should be encouraged in the future. Another development should focus on tracking multiple radio sources from different locations on a sky map simultaneously to derive both the global structure of the CME and exact location of the shock front. Especially if IPS systems are able to monitor radio sources over most of the period they are visible in the sky (as shown in Figures 3 and 9 from the LOFAR data) a better fitting of models to observations is possible. Such observations could be achieved by the next-generation IPS observation system of ISEE.

## Acknowledgements

This work was supported by MEXT/JSPS KAKENHI Grant Number 18H01266, 19K22028, and 21H04517. The observations of IPS were provided by the solar wind program of the Institute for Space-Earth Environmental Research (ISEE). This study was conducted using the computational resources of the Center for Integrated Data Science, ISEE, Nagoya University, through a joint research program. We thank the LASCO coronagraph group for the white-light CME images. This paper makes use of data obtained with the International LOFAR Telescope (ILT) under project code DDT8_006. LOFAR (van Haarlem et al., 2013) is the Low Frequency Array designed and constructed by ASTRON. Two of us (RAF and MMB) were partially supported by the LOFAR4SW project, funded by the European Community's Horizon 2020 Programme H2020 INFRADEV-2017-1 under grant agreement 777442. MMB was also supported in part by the STFC In-House Research grant to the Space Physics and Operations Division at UKRI STFC RAL Space.